\documentclass[12pt]{article}
\usepackage{amsmath}
\usepackage{latexsym}
\usepackage{amssymb}
\def\tr{\mathop{\rm tr}\nolimits}
\def\Id{\mathop{\rm I}}
\def\SE{\section}
\def\Label{\label}
\def\su{\mathfrak{su}}
\begin{document} %\draft
\vskip 2em 
\begin{center}
{\Large\bf $\su(1,1)$ coherent states and \par
a normal extension of $\su(1,1)$ annihilation operator:} \par
{\large\bf squeezed states and the simultaneous measurement \par
of $Q^{-1}P+PQ^{-1}$ and $Q^{-2}$ }
\vskip 1.5em
Fuminori SAKAGUCHI \par
Department of Electrical and Electronics Engineering \par
 Faculty of Engineering, Fukui University, 3-9-1 Bunkyo, 
Fukui-shi, 910-8507 Japan \par
 tel:+81-776-27-8912, fax:+81-776-27-8749, 
e-mail:saka@dignet.fuee.fukui-u.ac.jp\par
\vskip 1.5em
Masahito HAYASHI \par
Department of Mathematics, Faculty of Science \par
  Kyoto University, Kyoto, 606-8502 Japan \par
 tel:+81-75-753-3695, fax:+81-75-753-3711,
e-mail:masahito@kusm.kyoto-u.ac.jp \par
\end{center}
\begin{abstract}
The the over-complete eigenvector system of the operator $Q^{-1}P$ 
($Q$:position, $P$:momentum) which consists of the squeezed states 
$|0;\mu,\nu\rangle$ with various $\mu$ and $\nu$ are investigated 
from the viewpoint of the annihilation and creation relations related 
to the algebra $\su(1,1)$.  We derive a positive operator-valued measure(POVM) 
for the simultaneous measurement between the self-adjoint and 
anti-self-adjoint parts of $PQ^{-1}$.
\end{abstract}
\SE{Introduction}
As is well known, the eigenvectors of the linear combination of the 
boson annihilation and creation operators are the squeezed states[1], 
and the eigenvalue of the squeezed state 
 $|\alpha; \mu, \nu \rangle$ is a function both of the shift 
parameter $\alpha$ and of the squeezing parameters $\mu$ and $\nu$. 
In this eigenvalue problem, because the definition of the operator 
itself depends on the squeezing parameters $\mu$ and $\nu$ 
(i.e. the coefficients of linear combination are these parameters), 
we can not investigate the estimation problem and 
the uncertainty relation only with respect to the squeezing 
parameters $\mu$ and $\nu$.\par
In this paper, for investigating them only with respect 
to the squeezing parameters, we discuss another type 
of eigenvector problem related to the squeezed states.
The eigenvector system of the non-hermitian operator  $Q^{-1}P$ 
($Q$:position, $P$:momentum) is the set of the the squeezed states 
with $\alpha=0$, i.e. $\{~|0;\mu,\nu\rangle ~~|~~ \mu ,\nu {\rm :complex},~|\mu |^2- | \nu |^2=1 \}$.
The eigenvalue corresponding to the state $|0;\mu,\nu\rangle $ is
$i\frac{\mu +\nu }{\mu- \nu}$.  
Because states $|0;\mu,\nu\rangle $ and $|0;\mu',\nu'\rangle $ 
are the same states (with the phase difference neglected) 
in the case of
$i\frac{\mu +\nu }{\mu- \nu}=i\frac{\mu' +\nu' }{\mu'- \nu'}$, 
we may regard the eigenvalue $i\frac{\mu +\nu }{\mu- \nu}$
as the squeezing parameter in the case of $\alpha=0$.
Because $Q^{-1}P = (1/2)(Q^{-1}P + P Q^{-1}) -(i/2)Q^{-2}$,
the squeezed states with $\alpha=0$ are the minimum uncertainty states between 
$Q^{-1}P + P Q^{-1}$ and $Q^{-2}$.\par
These eigenvector relations are closely related to the Lie algebra $\su(1,1)$, 
because the three generators of the displacement of the squeezing 
parameters satisfy the commutation relations of this algebra. 
The above squeezed states $\{~|0;\mu,\nu\rangle ~~|~~ \mu ,\nu 
{\rm :complex},~|\mu |^2- | \nu |^2=1 \}$ are generalized 
coherent states associated with the Lie group generated 
by these generators.
Its action on the set of squeezed states 
$\{~|0;\mu,\nu\rangle ~~|~~ \mu ,\nu {\rm :complex},~
|\mu |^2- | \nu |^2=1 \}$ is covariant with
the natural action of $\su(1,1)$ on the left half plane of 
 the eigenvalue $i\frac{\mu +\nu }{\mu- \nu}$.
In addition, we can map this half plane into the unit circle
by a M\"{o}bius transformation.  By this mapping, 
we will derive the annihilation/creation relations and the number 
operators related to this algebra.  \par
We will derive these relations from the algebraic 
structure of the $\su(1,1)$ itself. The unitary representation 
of this algebra is related to the eigenvector problem 
of  $Q^{-1}P$, and this type of eigenvector problems can be commonly 
discussed for general  $\su(1,1)$. Moreover, we will show that the  
M\"{o}bius transform is 
corresponding to the change of the choice of the basis operators 
of the same algebra where one operator in the triplet is a 
linear function of the number operator.  By using this choice, 
we will investigate the annihilation and creation relations related to 
$\su(1,1)$. 
These relations are different from the boson 
annihilation and creation relations, because the product between the 
annihilation and creation operators is not a linear function but a 
kind of non-linear rational function of the number operator.  
From these relations, we will derive a non-linear type of 
reordering relation  between the 
annihilation and creation operators.
Moreover we derive the relations these annihilation and creation 
operators and $\su(1,1)$-coherent state.\par
Next we will discuss about the existence of a normal extension.
If the overcomplete eigenvector system of $A$ makes 
a pseudo-type (not a projection-type) of 'resolution of identity',
then we can derive the Positive Operator-Valued Measure (POVM) 
for the generalized measurement 
which is the optimal to measure
the non-hermitian operator $A$
(i.e. to simultaneously measure the operators
$(A + A^*)/2$ and $(A-A^*)/2$).
The general formalism proposed above for $\su(1,1)$ has a 
systematic method for this under a condition on the minimum 
eigenvalue of a basis operator.  This condition is not satisfied
 in the case of the eigenvalue problem of $Q^{-1}P$. However, 
we will show that it is satisfied in the case of the eigenvalue 
problem of $PQ^{-1}$.
\SE{Squeezed States and the eigenvalue problem with respect to the 
squeezing parameters}
Let $a_b:=\sqrt{1/2}(Q+iP)$ ($Q$:position, $P$:momentum) be the 
boson annihilation operator, and let 
\begin{equation*}
b_{\mu,\nu} := \mu a_b + \nu a_b ^*  ( |\mu |^2- | \nu |^2=1) .
%\Label{1}
\end{equation*}
Then, as is well known, this operator has complex eigenvalues and 
the eigenvector of $b_{\mu,\nu}$ associated with the eigenvalue $\alpha$ 
is the squeezed states vector $|\alpha; \mu, \nu \rangle$. The eigenvalue 
$\alpha$ indicates the center of the localization of the wave packet 
in the phase plane, while the coefficients $\mu$ and $\nu$ indicates 
the squeezing properties[1].  Therefore the set 
$\{~|0;\mu,\nu\rangle ~~|~~ \mu ,\nu {\rm :complex}$ is the 
set of the squeezed states located around the origin of the phase plane 
with various squeezing parameters.\par
From 
\begin{equation*}
b_{\mu,\nu} |0;\mu,\nu\rangle = 0, %\Label{2}
\end{equation*}
by operation $Q^{-1}$ from the left, we have
\begin{equation*}
(\mu + \nu) |0;\mu,\nu\rangle - i (\mu - \nu) Q^{-1}P |0;\mu,\nu\rangle = 0, 
%\Label{3}
\end{equation*} 
and hence
\begin{equation}
Q^{-1}P |0;\mu,\nu\rangle = i\frac{\mu +\nu }{\mu- \nu} |0;\mu,\nu\rangle .
\Label{4}\end{equation}
This relation is another kind of characteristic equation of the squeezed states.
The squeezed states with $\alpha=0$ can be regarded as the eigenvectors 
of the operator $Q^{-1}P$ associated with the eigenvalue $i\frac{\mu +\nu }{\mu- \nu}$.
This relation is very convenient for investigating the uncertainty relation 
and the quantum estimation problem only with respect to the squeezing parameters, 
because the operator  $Q^{-1}P$ itself does not but the eigenvalue does depend 
on $\mu$ and $\nu$.\par
The operator $Q^{-1}P$ is not self-adjoint, and the self-adjoint part and 
the anti-self-adjoint part of $Q^{-1}P$ are $(Q^{-1}P + P Q^{-1})/2$ and 
$Q^{-2}/2$, respectively.  
Therefore $|0;\mu,\nu\rangle$ is a minimum-uncertainty state between
$Q^{-1}P + P Q^{-1}$ and $Q^{-2}$ in the sense that the equality in 
the uncertainty relation\footnote{the inequality (\ref{5}) is concerning with
the width of wave packet, not but measuring error. }
\begin{equation}
\bigl(\Delta(Q^{-1}P + P Q^{-1})\bigr)^2 \cdot \bigl(\Delta(Q^{-2})\bigr)^2 
= \frac{1}{4} ~\Bigl| i \langle\psi | [Q^{-1}P + P Q^{-1}, Q^{-2}] | \psi \rangle \Bigr|^2 
\Label{5}\end{equation}
holds.\par
On the other hands, similar relations are shown for the eigenvectors 
of the operator $PQ^{-1}$.  In this case, the eigenvectors are not 
the usual squeezed states.  However, in Sec.4, we will show that 
they are the vectors obtained by squeezing the one-boson state.  
Moreover, because $PQ^{-1}$ is the adjoint of ${Q^{-1}P}$, these vectors are 
the minimum uncertainty states between the same pair $(Q^{-1}P + P Q^{-1})$ 
and $Q^{-2}$. \par
The eigenvalue problem $Q^{-1}P$ and $PQ^{-1}$ has the algebraic 
structures discussed 
in the following sections.  In the next section, we will start with more 
general algebraic formalism.
\SE{Annihilation and Creation Relations Related to $\su(1,1)$ and the 
corresponding $\su(1,1)$ coherent states}
The triplet of the anti-self-adjoint operators $E_0$, $E_+$ and $E_-$ 
on a Hilbert space ${\cal H}$ which satisfy the commutation relations,  
\begin{equation*} 
[ E_0, E_\pm ] = \pm 2 E_\pm,  \quad [ E_+, E_- ] = E_0 
%\Label{6}
\end{equation*}
is called the unitary representation of the Lie algebra $\su(1,1)$. 
This algebra is isomorphic to the Lie algebra $\mathfrak{sl}(2,\mathbb{R})$. 
From them, define another triplet of the operators $L_0$, $L_+$ and $L_-$ by 
\begin{equation}
L_0 := i(E_- - E_+), \quad 
L_\pm := \frac{1}{2}\bigl(E_0 \pm i(E_+ + E_-)\bigr).
\Label{7}\end{equation}
Then the same type of commutation relations 
\begin{equation} 
[ L_0, L_\pm ] = \pm 2 L_\pm,  \quad [ L_+, L_- ] = L_0 
\Label{8}\end{equation}
hold, and these are another basis of the same Lie algebra. However, 
in this basis system, $L_0$ is self-adjoint while $L_\pm$ are 
neither self-adjoint nor anti-self-adjoint, and 
$(L_\pm)^* = -L_\mp$.
From (\ref{7}), $E_0$, $E_+$ and $E_-$ are written 
by $L_0$, $L_+$ and $L_-$, as 
\begin{equation}
E_0 = L_+ + L_-, \quad E_\pm = \pm \frac{i}{2} (L_0 \mp L_+ \pm L_-).
\Label{9}\end{equation}
In this paper, we investigate only the cases where the representation 
is irreducible and is not trivial.
Then the corresponding Casimir operator which should be a scalar 
by the Schur's lemma, 
\begin{equation}
C := L_0^2 + 2 (L_+ L_- + L\lineskip .5em_- L_+) = \beta 
\Label{10}\end{equation}
where the parameter $\beta$ depends on the representation of the Lie algebra. 
From (\ref{8}), this relation can be written in other forms
\begin{equation}
 L_0^2 + 2 L_0 + 4 L_- L_+ = \beta, \quad L_0^2 - 2 L_0 + 4 L_+ L_- = \beta. 
\Label{11}\end{equation}
From (\ref{7}) and (\ref{10}),
\begin{equation}
 E_0^2 + 2 E_0 + 4 E_- E_+ = \beta, \quad E_0^2 - 2 E_0 + 4 E_+ E_- = \beta. 
\Label{13}
\end{equation}
From the commutation relation relations (\ref{8}), 
we can show that if $v$ is the eigenvector of $L_0$ associated with the 
eigenvalue value $\kappa$ then $\kappa +2$ is 
also its eigenvalue, by  
\begin{align}
L_0 (L_+ v) &= L_+ (L_0 + 2) v = (\kappa +2)(L_+ v) 
\nonumber \\
L_0 (L_- v) &= L_- (L_0 - 2) v = (\kappa -2)(L_- v) .
\Label{15}\end{align}
From this relation and the self-adjoint property of $L_0$, that 
the eigenvector system of $L_0$ is a orthogonal system and the 
eigenvalues of $L_0$ is uniformly spaced real numbers. From the 
irreducibility and the unitarity, it is easily shown 
that the dimension of the kernel of  $L_\pm$ should be not more than 
one and it is impossible both of ${\rm dim}~ {\rm Ker}~ L_\pm$ are 
one simultaneously.
From now, we are investigating the case where ${\rm dim}~ {\rm Ker}~ L_+ = 0$ 
and ${\rm dim}~ {\rm Ker}~ L_- = 1$. The opposite case can be reduced to 
this case by the change of the definition of $L_0$ and $L_\pm$ without 
loss of generality(We will not investigate the case where both of 
${\rm dim}~ {\rm Ker}~ L_\pm$ are zero). 
Let $v_0$ be the unit vector in ${\rm Ker}~ L_-$, 
from the relation (\ref{15}), $v_0$ should be the eigenvector of $L_0$ associated with 
the minimum eigenvalue $\lambda$ (otherwise the existence of a 
smaller eigenvalue were contradictory to $L_- v_0 =0$).  
Then the characteristic equation 
\begin{equation}
L_0 \bigl( (L_+)^n v_0 \bigr) = (\lambda + 2n) \bigl( (L_+)^n v_0 \bigr) 
\Label{16}\end{equation}
holds. 
It is well known that this constant $\lambda$ uniquely determine the 
representation of the algebra $\su(1,1)$.
In this case, the unitarity means that $\lambda \,> 0$ [6].
Thus this representation space determined by the constant $\lambda$ is
denoted by ${\cal H}_{\lambda}$.
From the irreducibility and the self-adjointness of $L_0$,
we can show easily that 
$\{ (L_+)^n v_0 \}_{n=0}^{\infty}$ is a CONS of ${\cal H}$.
%which consists of the eigenvalues of $L_0$. 
From the relation(\ref{11}) 
and $L_- v_0 =0$ 
and $L_0 v_0 = \lambda v_0$, we have 
\begin{equation}
\beta = \lambda(\lambda-2)
\Label{17}\end{equation}
(NB: Similar relations does not hold for $E_0$ and $E_+$, because 
$E_0$ is anti-self-adjoint.) \par
Now, we define the $\su(1,1)$ number operator $N$ and 
$\su(1,1)$ number states vector $|n\rangle_{N}$ by 
\begin{align}
N &:= \frac{1}{2}(L_0 - \lambda) \Label{44} \\
|n\rangle_N &:= \sqrt{\frac{\Gamma (\lambda )}{n!~\Gamma (\lambda+n )}} 
L_+^n v_0, \Label{46}
\end{align}
then, from (\ref{16}) we have $
N |n\rangle_N = n |n\rangle_N $,
where $| x \rangle_X$ means the unit eigen vector of an operator $X$
associated with an eigenvalue $x$.
We can show the right hand side of (\ref{46}) is an unit vector
from the relation 
\begin{align*}
&\langle L_+^n v_0, L_+^n v_0\rangle 
= - \langle L_+^{n-1} v_0, (L_- L_+) L_+^{n-1} v_0\rangle \\ 
=& -\frac{1}{4} \langle L_+^{n-1} v_0, (\beta-L_0^2-2L_0)) 
L_+^{n-1} v_0\rangle 
= n(\lambda +n-1) \langle L_+^{n-1} v_0, L_+^{n-1} v_0\rangle 
\end{align*}
where we utilize (\ref{11}) with (\ref{17}).\par
Next we will define the $\su(1,1)$ annihilation operator $a$.
From the relation (\ref{16}),
we can prove that $(L_0 - \lambda)| n \rangle_N$ belongs to  
the range of $L_+$ for any $n$.
Because $\dim {\rm Ker}(L_+)= 0$,
we can define $\frac{1}{2} L_+^{-1} (L_0 - \lambda) | n \rangle_N$.
Therefore, we can define that
the $\su(1,1)$ annihilation operator
\begin{equation}
a: = \frac{1}{2} L_+^{-1} (L_0 - \lambda).
\Label{18}\end{equation}
on the dense subset $\left\{ \sum_{n=0}^{\infty} x_n | n \rangle_N 
\in {\cal H}
\left| 
\sum_{m=0}^{\infty}
\left|
\sum_{n=0}^{\infty}
x_n ~_N\langle m |, \frac{1}{2} L_+^{-1} (L_0 - \lambda) | n \rangle_N
\right|^2 \,< \infty \right. \right\}$.
With the unitary displacement operator 
\begin{equation}
D(\xi) := \exp \left( \xi L_+ - \bar{\xi} L_+^* \right) ,
\Label{19}\end{equation}
define the $\su(1,1)$ coherent state 
\begin{align}
v(\zeta) &: = D\left(\frac{1}{2}e^{i\arg \zeta}\ln\frac{1+|\zeta|}{1-|\zeta|}
\right)~| 0 \rangle_n \nonumber \\ 
&= \exp(\zeta L_+) \exp \left( \frac{1}{2} \ln (1-|\zeta|^2) ~L_0 \right) 
\exp(\bar{\zeta} L_-) ~| 0 \rangle_N, \quad (|\zeta|<1).
\Label{20}\end{align}
(The latter ``normal-order'' form of the right hand side is 
obtained from the relation 
given in pp.73-74 of Ref[2], with the correspondences 
$K_0=L_0/2$, $K_+=L_+$ and $K_-=-L_-$.)  
Then, from $K_- | 0 \rangle_N = 0$ and 
$K_0 | 0 \rangle_N =(\lambda/2) | 0 \rangle_N$,
\begin{equation}
\exp\left( \frac{1}{2} \ln (1-|\zeta|^2) ~L_0\right) \exp(\bar{\zeta} L_-) ~
| 0 \rangle_N
= (1-|\zeta|^2)^{\lambda /2} | 0 \rangle_N.
\Label{21}\end{equation}
Because $[ a, L_+ ] = 1/2 L_+^{-1} [L_0,L_+] = 1$, 
\begin{equation}
\left[ a  ,\exp(\zeta L_+) \right] = \zeta \exp(\zeta L_+), 
\Label{22}\end{equation}
hence, from the relations (\ref{18})-(\ref{22}) and 
$(L_0 - \lambda)| 0 \rangle_N =0$, 
we have
\begin{equation}
a v(\zeta) = \exp(\zeta L_+)~ a ~| 0 \rangle_N + \zeta \exp(\zeta L_+) 
| 0 \rangle_N 
= \zeta v(\zeta).
\Label{23}\end{equation}
These relations show that the vector $v(\zeta)$ is the eigenvector 
of $a$ associated with the eigenvalue $\zeta$. 
Thus we can denote $v(\zeta)$ by $| \zeta \rangle_{a}$.\par
Next, we investigate the annihilation and creation relations. 
Then, from the (\ref{8}) and (\ref{18}), 
\begin{equation*}
[a, N] = \frac{1}{4} [L_+^{-1}, L_0] (L_0 - \lambda) = \frac{1}{2} 
L_+^{-1} (L_0 - \lambda) =  a ,
%\Label{45}
\end{equation*}
this implies that the operator $a$ is the annihilation operator 
of the eigenvector system of $N$.
Because we get 
$a L_+^n |0 \rangle_n = n L_+^{n-1} | 0 \rangle_N$ from 
(\ref{16}) and (\ref{18}),
by (\ref{46}), we have the annihilation and creation relations 
\begin{equation}
a |n\rangle_N = \sqrt{\frac{n}{n+\lambda -1}} ~|n-1\rangle_N 
\quad {\rm and}\quad
 a^* |n\rangle_N = \sqrt{\frac{n+1}{n+\lambda}} ~|n+1\rangle_N,
\Label{51}\end{equation}
where the first equation of (\ref{51}) means that $a|0 \rangle_{N}= 0$
in the case of $\lambda=1, n=0$.
Therefore, from the completeness and orthogonality of the eigenvectors 
of $L_0$,
we get the following relations in the case of $\lambda \neq 1$
\begin{equation*}
a^* a = (N + \lambda -1)^{-1} N \quad {\rm and} \quad
a a^* = (N + \lambda)^{-1} ( N + 1 ).
%\Label{52}
\end{equation*}
By eliminating $N$ from these, we have the re-ordering relation 
between the annihilation $a$ and the creator $a^*$,
\begin{equation}
a a^* = - ( a^* a + \lambda -2 )^{-1} ( \lambda a^* a -1 ) 
\quad{\rm and} \quad
a^* a = ( a a^* - \lambda )^{-1} \bigl( (2-\lambda) a a^* -1 \bigr).
\Label{53}\end{equation}
In the case of $\lambda=1$, we have $a a^* = 1 , \quad a^* a = 1 - 
| 0 \rangle_N ~_N\langle 0 |$.
These relations are important for the calculating
the quantum characteristic function.\par
%(The vector $v(\zeta)$ 
%is the generalized coherent states related to $\su(1,1)$.)  
%Because of the unitarity of the displacement operator, 
%$v(\zeta)$ is normalized with normalized $v_0$.
From the relations (\ref{8}), (\ref{9}), (\ref{11}) with (\ref{17}) and the relation 
$\left[L_0,L_+^{-1} \right]=L_+^{-1}[L_+,L_0]L_+^{-1}=-2L_+^{-1}$,
we can show the following relation
\begin{align*}
2 (E_0 - \lambda) (a-1) L_+ 
=& (L_+ + L_- - \lambda) \bigl(L_0 - (\lambda -2) - 2 L_+\bigr) \\ 
=& \lambda (\lambda-2) + \lambda ( -L_+ - L_- - L_0 + 2 L_+) \\
&\quad + ( L_+ L_0 + L_- L_0 - 2 L_+ L_+ - 2 L_- L_+ + 2 L_+ + 2 L_-) \\ 
=& - \lambda (L_0 - L_+ + L_-) + L_0 L_0 + 2 L_0 + 2 L_- L_+  \\
&\quad + (2 L_0 L_+ - 4 L_+ - L_+ L_0) + L_- L_0 - 2 L_+ L_+ + 2 L_+ + 2 L_-\\ 
=& (L_0 - L_+ + L_-) ( -\lambda + L_0 + 2 + 2 L_+)\\
=& -2i E_+ \bigl(L_0 - (\lambda -2) + 2 L_+\bigr) = -4i E_+ (a+1) L_+  \\
(E_0 - \lambda)(a-1) | 0 \rangle_N
= &- (E_0 - \lambda) | 0 \rangle_N
=(L_0 - L_+ ) | 0 \rangle_N
= -2i E_+ | 0 \rangle_N
= -2i E_+ (a+1) | 0 \rangle_N .
%\Label{24}
\end{align*}
Hence we get
\begin{equation} 
(E_0 - \lambda) ( a - 1 ) = -2i E_+ ( a + 1 ).
\Label{25}\end{equation}
If the dimension of the kernel of $E_+$ is not 0,
then this representation becomes trivial by the relation (\ref{13})
and the unitarity. 
Using the relation (\ref{51}) and (\ref{25}), we can show that
$(E_0 - \lambda)|n \rangle_N$ belongs to the range of $E_+$.
Therefore, we can define the operator $A$ like the case of $a$
\begin{equation}
A := \frac{1}{2} E_+^{-1} (E_0 - \lambda).
\Label{26}\end{equation}
Hence, we have 
\begin{align}
A =& -i (a + 1) ( a - 1 )^{-1}
\Label{27} \\
A | \zeta \rangle_{a} =& -i \frac{\zeta +1}{\zeta -1} | \zeta \rangle_{a}.
\Label{28}\end{align}
Thus the vector $| \zeta \rangle_{a} $ can be denoted by
$\displaystyle{\left| - i \frac{\zeta+1}{\zeta-1} \right \rangle_{A}}$.
Since the imaginary number $-i$ doesn't belong to the spectrum of $A$,
the relation (\ref{27}) indicates that 
\begin{equation*}
a = (A + i)^{-1} (A - i) .
%\Label{29}
\end{equation*}
The operator $A$ has a similar property to (\ref{23}). Define 
the `normal-ordered affine coherent vector' as
\begin{equation}
{v'}_n(s,t): = \exp(sE_+) \exp(tE_0) | 0 \rangle_n . 
\Label{30}\end{equation}
Then, because $[ A,E_+ ] = (1/2) E_+^{-1} [E_0,E_+] = 1$, 
we have 
\begin{align}
A ~\exp(sE_+) =& \exp(sE_+)~ A + s \exp(sE_+) \Label{31} \\
A ~\exp(tE_0) =& e^{2t}\exp(tE_0)~ A. \Label{32}
\end{align}
From the relations (\ref{26}),(\ref{30})-(\ref{32}) 
and (\ref{28}) with $\zeta=0$, 
\begin{equation*}
A {v'}_n(s,t) =e^{2t}\exp(sE_+)\exp(tE_0)~ A ~| 0 \rangle_n 
+ s \exp(sE_+)\exp(tE_0) | 0 \rangle_n 
= (e^{2t}i+s) {v'}_n(s,t). %\Label{33}
\end{equation*}
We have similar relations under other orderings. Define 
the `anti-normal-ordered affine coherent vector' as
\begin{equation}
{v'}_a(s,t) := \exp(tE_0)\exp(sE_+) | 0 \rangle_n. 
\Label{34}\end{equation}
Then, from (\ref{26}),(\ref{31}),(\ref{32}),(\ref{34}) and (\ref{28}) 
with $\zeta=0 $, 
\begin{equation}
A {v'}_a(s,t) 
=e^{2t}\exp(tE_0)\exp(sE_+)~ A ~| 0 \rangle_n 
+ s e^{2t} \exp(tE_0)\exp(sE_+) | 0 \rangle_n
= e^{2t}(i+s) {v'}_a(s,t).
\Label{35}\end{equation}
These relations show that the vectors ${v'}_n(s,t)$ and ${v'}_a(s,t)$ 
are the eigenvectors of $A$ associated with the eigenvalue 
$e^{2t}i+s$ and $e^{2t}(i+s)$, respectively. 
Thus the vectors ${v'}_n( {\rm Re}~ \eta , 1/2 \ln {\rm Im}~\eta )$ 
and ${v'}_a( ({\rm Re}~\eta)/({\rm Im}~\eta) ,1/2 \ln {\rm Im}~\eta )$ 
are equivalent to $| \eta \rangle_{A}$ with the phase difference neglected.
Note that the operators $\exp(sE_+)$ and  $\exp(tE_0)$
are unitary because of $E_0$ and $E_+$ are anti-self-adjoint.

From the viewpoint of the theory of $\su(1,1)$ coherent states[2], 
when $\lambda > 1$, these relations implies that the resolutions 
of unity 
\begin{align}
\int_{\rm U} |\zeta\rangle_a ~_a\langle \zeta| \mu_\lambda(\,d\zeta) &= \Id
\quad \hbox{ with } \quad 
\mu_\lambda(\,d\zeta) 
:= \frac{\lambda-1}{\pi}\frac{\,d^2\zeta}{(1-|\zeta |^2)^2}
\Label{40} \\
\int_{\rm H} |\eta\rangle_A ~_A\langle \eta| \nu_\lambda(\,d\eta) &= \Id
\quad \hbox{ with } \quad 
\nu_\lambda(\,d\eta) 
:= \frac{\lambda-1}{4\pi}\frac{\,d^2\eta}{({\rm Im}~\eta)^2},
\nonumber %\Label{41}
\end{align}
where U (H)  denotes the inside of the unit circle (the upper half plane),
respectively.
However, when $\lambda \,< 1$, these relations are impossible. 

\SE{Squeezed States as the Eigenstates of $Q^{-1}P$ and \\
'Odd Squeezed States' as those of $PQ^{-1}$} 

In this section, we will interpret the eigenfunction problem (\ref{4}) 
from the algebraic structure discussed in Sec. 3. For this, 
we have only to choose the representation 
\begin{equation*}
E_0 = i (PQ+QP)/2, \quad E_+ = i Q^2/2, \quad E_- = - i P^2/2. 
%\Label{54}
\end{equation*}
Then from (\ref{7})
we get the following relations
\begin{equation}
L_0 = n_b +\frac{1}{2}, \quad L_+ = -(1/2) {a_b^*}^2, \quad 
L_- = (1/2) a_b^2,
\Label{55}\end{equation}
where $n_b:=1/2(Q^2+P^2-1)= a_b a_b^*$ 
is the boson number operator and $a_b$ 
is the boson annihilation operator defined in Sec. 2.
The Casimir operator $C$ is represented as
\begin{equation*} 
\beta = C = - (PQ+QP)^2/4 + (Q^2P^2+P^2Q^2)/2 = - \frac{3}{4}.
%\Label{56}
\end{equation*}
From (\ref{17}) we have $\lambda=1/2$ or $3/2$.  
These two solutions are corresponding 
to the two function spaces of the representation.  
When we choose 
the function space with even parity
\begin{equation*}
L^2_{\rm even} := \{ f(q)\in L^2(R) | f(-q)=f(q) \} , 
%\Label{57}
\end{equation*}
then $\lambda=1/2$. 
Therefore, there is no POVM defined in (\ref{40}).
From (\ref{18}),(\ref{26}),(\ref{44}) and (\ref{55}),
we have
\begin{align*}
A =& \frac{1}{2} Q^{-2} (PQ+QP+i) 
= Q^{-2} QP = Q^{-1}P ,
%\Label{58} 
\\
a =& - {a_b^*}^{-2} n_b = {a_b^*}^{-2} a_b^* a_b = {a_b^*}^{-1} a_b 
%\Label{59} 
\\
N =& \frac{1}{4}(Q^2+P^2-1) = \frac{1}{2} n_b.
%\Label{60}
\end{align*}
Then, from (\ref{4}), (\ref{46}) and (\ref{55}),  
\begin{align*}
|0;\mu,\nu\rangle = \left|i\frac{\mu+\nu}{\mu-\nu}\right\rangle_A ,
\quad %\hbox{ and } \quad
|n\rangle_N = (-1)^n |2n\rangle_{n_b},
%\Label{62}
\end{align*}
where $| n \rangle_{n_b}$ denotes the boson number state.
From (\ref{51}) and (\ref{53}), we have the annihilation and 
creation relations 
\begin{equation*}
a |n\rangle_N = \sqrt{\frac{2n}{2n-1}} ~|n-1\rangle_N,
\quad a^* |n\rangle_N =  \sqrt{\frac{2n+2}{2n+1}} ~|n+1\rangle_N 
%\Label{63}
\end{equation*}
and the non-linear re-ordering relations 
\begin{equation*}
a a^* = - ( 2a^* a -3 )^{-1} ( a^* a -2 ) \quad{\rm and} \quad
a^* a = ( 2a a^* - 1 )^{-1} ( 3 a a^* -2 ) .
%\Label{64}
\end{equation*}

On the other hand, when we choose 
the function space with odd parity
\begin{equation*}
L^2_{\rm odd} := \{ f(q)\in L^2(R) | f(-q)=-f(q) \} , 
%\Label{65}
\end{equation*}
then $\lambda=3/2$. 
There is the POVM defined as (\ref{40}).
We have
\begin{align}
A =& \frac{1}{2} Q^{-2} (PQ+QP+3i) 
= Q^{-2} (QP+i) = PQ^{-1} ,
\Label{66}\\
a =& - {a_b^*}^{-2} (n_b - 1) 
= {a_b^*}^{-2} (a_b^* a_b - 1) = a_b {a_b^*}^{-1} 
%\Label{67} 
\nonumber \\
N =& \frac{1}{4}(Q^2+P^2-3) = \frac{1}{2} (n_b - 1)
\nonumber %\Label{68}
\end{align}
and hence, from (\ref{46}) and (\ref{55}), 
\begin{equation}
|n\rangle_N = (-1)^n |2n+1\rangle_{n_b}.
\Label{69}\end{equation}
In this case the $\su(1,1)$ coherent state is the eigen state
of the `squeezed number' operator $D(\xi) n_b D(\xi)^* $ associated with
the eigenvalue 1.
From the relations (\ref{30}), (\ref{44}), (\ref{66}) and (\ref{69}) 
with $n=0$,
we can show that the eigenvector of $PQ^{-1}$ is obtained 
by squeezing the one-boson state, as
\begin{equation}
|2p+e^{4q}i\rangle_{PQ^{-1}} = \exp (ipQ^2) \exp \bigl(iq(PQ+QP)\bigr) 
~|1\rangle_{n_b}.
\Label{70}\end{equation}
Hence we call it the 'odd squeezed state' for convenience in this paper.
In this 'odd' case, the annihilation and creation relations 
and the non-linear re-ordering relations are 
\begin{align*}
a |n\rangle_N = \sqrt{\frac{2n}{2n+1}} ~|n-1\rangle_N 
\quad &{\rm and } \quad 
a^* |n\rangle_N =  \sqrt{\frac{2n+2}{2n+3}} ~|n+1\rangle_N 
%\Label{71} 
\\
a a^* = - ( 2a^* a -1 )^{-1} ( 3a^* a -2 ) \quad &{\rm and}  \quad
a^* a = ( 2a a^* - 3 )^{-1} ( a a^* -2 ) .
%\Label{72}
\end{align*}
\SE{Relation to the Cauchy Wavelets}
It is known that the system of the wavefunctions of the 
eigenvectors of the operator $Q-ikP^{-1}$ ($k$: positive real number) 
is the wavelet system[3] of the Cauchy wavelets i.e. 
$\{\sqrt{\frac{1}{|a|}}h_k(\frac{q-b}{a})~|~(a,b)\in 
{\rm (upper-half~ plane)}\}$ with 
$h_k(q):=\frac{({\rm const.})}{(q+i)^{k+1}}$ [4,5].
This eigenvalue problem also has the same algebraic structure 
discussed in the previous sections.  In this case, we choose 
\begin{equation}
E_0 = -i (PQ+QP), ~~~ E_+ = i P, ~~~ E_- = -i (QPQ+k^2P^{-1}), 
\Label{73}\end{equation}
and then 
\begin{align*}
L_0 &= (1/2) \bigl( QPQ +k^2P^{-1} + P \bigr),  \\ 
L_\pm &= \pm(i/2) \bigl( QPQ \mp (PQ+QP) +k^2P^{-1} - P \bigr), 
%\Label{74}
\end{align*}
and $\beta = 4k^2-1$.  
Now we choose 
the function space 
$L^2(\mathbb{R}^+)$ in the momentum representation, which is 
irreducible.
The vector $H_k$ 
defined in (\ref{77}) belongs to the kernel of 
$L_-$
\begin{equation}
H_k(p):=
\frac{1}{\sqrt{2^{2k+1}\Gamma(2k+1)}}
\cdot p^k e^{-p}   .
\Label{77}\end{equation}
From a simple calculation, the eigen value of $L_0$ for $H_k$ is
$\lambda = 2k+1$. And we get
\begin{align*}
A =& - P^{-1}\bigl(PQ+QP-(2k+1)i\bigr) = -2 (Q-ikP^{-1})
%\Label{75}
\\
N =& (QPQ+k^2P^{-1}+P-2k-1)/2.
%\Label{76}
\end{align*}
In the position representation, the corresponding 
function space is 
$\{f(q)\in L^2~|~f_H(q)=if(q)\} (f_H:{\rm Hilbert~tansf.~ of~}f)$ 
or the space of normalizable analytic signals in signal analysis, 
and the eigenfunction of the vacuum vector is the basic wavelet 
$h_k(q):=\frac{({\rm const.})}{(q+i)^{k+1}}$ .
For these, we have the similar annihilation and creation relations 
and the non-linear re-ordering relations[5]
\begin{align*}
a |n\rangle_N &= \sqrt{\frac{n}{n+2k}} ~|n-1\rangle_N ~~~~ {\rm and}~~~ a^* |n\rangle_N = \sqrt{\frac{n+1}{n+2k+1}} ~|n+1\rangle_N 
%\Label{78}
\\
a a^* &= - \bigl( a^* a + (2k-1) \bigr)^{-1} \bigl( (2k+1)a^* a - 1 \bigr)
%\Label{79}
\\
a^* a &= - \bigl( a a^* + (2k+1) \bigr)^{-1} \bigl( (2k-1)a a^* - 1 \bigr).
%\Label{80}
\end{align*}
\SE{Normal extension of the annihilation operator $a$ and 
POVM for simultaneous measurement of $PQ^{-1}+PQ^{-1}$ and $Q^{-2}$}
In this section, we consider a normal extension of $\su(1,1)$ 
annihilation operator $a$.
In this case, It is sufficient to construct a normal extension of $A$.
Let $Z$ be a operator on ${\cal H}$ satisfying
\begin{equation}
[ Z , Z^* ] \ge 0  . \Label{200}
\end{equation}
If a triplet of a Hilbert space ${\cal H}'$, a state $\phi \in {\cal H}'$
and a normal operator $\tilde{Z}$ on ${\cal H} \otimes{\cal H}'$
satisfies the following relations
(\ref{201}) and (\ref{202}) for any density operator $\rho$ on ${\cal H}$,
the triplet $({\cal H}',\phi,\tilde{Z})$ is called a normal extension of 
the operator $Z$
\begin{align}
\tr \tilde{Z} \tilde{Z}^* \rho\otimes |\phi \rangle \langle \phi| &=
\tr Z Z^* \rho \Label{201}\\
\tr \tilde{Z} \rho \otimes |\phi \rangle \langle \phi| &=
\tr Z  \rho  \Label{202} .
\end{align}
Not satisfying the condition (\ref{200}),
there is no normal extension of $Z$.
But, the converse is not true.
Now, we consider the relation between 
a normal extension of $Z$ and a simultaneous measurement of
$X$ and $Y$, where $X$ and $Y$ are self-adjoint operators such that 
$Z=X+iY$.\par
If a POVM $M$ satisfies the condition (\ref{206}),
the POVM $M$ is called a simultaneous measurement
between $X$ and $Y$ [8]
\begin{equation}
Z = \int_{\mathbb{C}} z M (\,d z). \Label{206}
\end{equation}
If a POVM $M$ satisfies the condition (\ref{206}),
then we have the lower bound of the second moment for any 
density operator $\rho$ as
\begin{equation}
\int_{\mathbb{C}} |z|^2 \tr 
\rho M (\,d z) \ge \tr \rho Z Z^* .
\Label{105}\end{equation}
The inequality (\ref{105}) is derived by the following.
Using the relation (\ref{206}), we have 
\begin{equation}
0 \le \int_{\mathbb{C}}
(Z - z ) M(\,d z ) (Z - z)^* 
= \int_{\mathbb{C}} | z |^2
M(\,d z) - Z Z^*.\Label{207}
\end{equation}
The inequality (\ref{105}) can be derived from the inequality (\ref{207}).
Let $E_{\tilde{Z}}$ be the spectral measure of $Z$.
A POVM (Positive Operator-Valued Measure)
$M_{\tilde{Z}}$ on ${\cal H}$ is defined as
\begin{equation}
M_{\tilde{Z}}(\,d z) := \tr_{{\cal H}'} \Id \otimes | \phi \rangle
\langle \phi| E_{\tilde{Z}}(\,d z) ,
\end{equation}
where $\tr_{{\cal H}'}$ denotes the partial trace with respect to ${\cal H}'$.
From (\ref{201}) and (\ref{202}), the POVM $M_{\tilde{Z}}(\,d z)$
satisfies the condition (\ref{206}), and 
attains the lower bound of (\ref{105}).
Therefore it is the optimal simultaneous measurement of $X$ and $Y$ 
in the sense of the second moment.
Conversely,
It can be shown that if there is a POVM attaining the lower bound of 
(\ref{105}), there is a normal extension.\par
Next, the commutater of $A$ and $A^*$ is calculated as
\begin{align*}
[A, A^*] =& \frac{1}{4} \left[ E_+^{-1}(E_0-\lambda)
,(E_0+\lambda)E_+^{-1}\right] \\ 
=& \frac{1}{4} E_+^{-1} (E_0+\lambda) \left[ E_0,E_+^{-1} \right] 
+ \frac{1}{4} \left[  E_+^{-1} ,E_0 \right] E_+^{-1} (E_0-\lambda) \\ 
=& \frac{1}{2} E_+^{-1} \left[  E_+^{-1},E_0 \right] - \lambda E_+^{-2} 
= -(\lambda-1 ) E_+^{-2}= (\lambda-1 ) E_+^{-1} (E_+^{-1})^*.
%\Label{84}
\end{align*}
Since $E_+^{-1} (E_+^{-1})^* \ge 0$,
there may be a normal extension of $A$ in the case of $\lambda \,> 1$, and
there may be one of $A^*$ in the case of $0\,< \lambda \,<1$.
Remark that $E_+^{-1}(E_0 - 1)$ is a symmetric operator.
%self-adjoint operator.
Now, we construct a normal extension $({\cal H}',\phi,\tilde{A})$
of $A$ in the case of $\lambda \,> 1$ as the following method.
Let $\tilde{A}$ defined as
$\tilde{A}:=
A \otimes \Id' + E_+^{-1} \otimes F'$.
Since 
$[A, E_+^{-1}] = (1/2) E_+^{-1} [E_0, E_+^{-1}] 
= - E_+^{-2}$, we have 
\begin{align}
\left[\tilde{A}, \tilde{A}^* \right] 
= &[A, A^* ] \otimes \Id' 
- [ A , E_+^{-1}] \otimes {F'}^* 
+ [ E_+^{-1}, A^*] \otimes F'
- E_+^{-2} \otimes [ F' , {F'}^*] \nonumber \\
=& E_+^{-2} \otimes \left(
- (\lambda -1) + F' + {F'}^* - [ F', {F'}^*] \right) . \Label{86}
\end{align}
From (\ref{86}), the normality condition of $\tilde{A}$ means
\begin{equation*}
\left[ {F'}-{F'}^* , \frac{-i}{2} \left( {F'}^*+ {F'} - 
(\lambda-1)\right)\right]
=2  \left(
\frac{-i}{2} \left( {F'}^*+ {F'} - (\lambda-1)\right)\right). %\Label{87}
\end{equation*}
Remark that the operators 
${F'}^*-{F'}$ and $\frac{-i}{2} \left( {F'}^*+ {F'} - (\lambda-1)\right)$
are anti-self-adjoint.
This relation is the commutation relation of $E_0$ and $E_+$ of the
Lie algebra $\su(1,1)$.
Let a triplet $E_0'$, $E_+'$ and $E_-'$ be a unitary and irreducible 
representation of $\su(1,1)$ on ${\cal H}'$.
Choosing 
\begin{equation*}
{F'}-{F'}^* =  {E'}_0  \quad{\rm and}  \quad 
\frac{-i}{2} \left( {F'}^*+ {F'} - (\lambda-1)\right) = {E'}_+ ,
%\Label{90}
\end{equation*}
$\tilde{A}$ satisfies the normality condition.
Let
\begin{equation*}
L'_0 := i(E'_- - E'_+), 
\quad L'_\pm := \frac{1}{2}\bigl(E'_0 \pm i(E'_+ + E'_-)\bigr),
%\Label{81}
\end{equation*}
then the operators ${F'}$ and ${F'}^*$ are calculated as
\begin{equation*}
{F'} = -\frac{1}{2} \left(L'_0 - (\lambda-1)\right) + L'_+ , \quad
{F'}^* = -\frac{1}{2} \left(L'_0 - (\lambda-1)\right) - L'_- .
%\Label{91}
\end{equation*}
Next we consider the conditions (\ref{201}) and (\ref{202}).
The conditions (\ref{201}) and (\ref{202}) mean that
\begin{align}
\langle \phi ,{F'}^* \phi \rangle 
\tr A^* E_+^{-1} \rho 
- \langle \phi ,{F'} \phi \rangle 
\tr E_+^{-1} A \rho 
- \langle \phi, {F'} {F'}^* \phi \rangle \tr E_+^{-2} \rho 
&=0, \Label{203} \\
\langle \phi ,{F'} \phi \rangle 
\tr E_+^{-1} \rho &=0. \Label{204}
\end{align}
If ${F'}^* \phi = 0$, then 
the conditions (\ref{203}) and (\ref{204}) are satisfied.
If the lowest eigenvalue of $L'_0$ is 
$\lambda-1$, then
the $\su(1,1)$ vacuum state $|0 \rangle_{N'}$ belongs to the kernel of $F$.
Therefore we proved that 
the triplet $\left({\cal H}_{\lambda-1},|0 \rangle_{N'},
A \otimes \Id'+ E_+^{-1} \otimes \left( 
- \frac{1}{2} \left(L'_0 - (\lambda-1)\right) + L'_+\right)\right)$
is a normal extension of $A$.
In addition, the POVM $| \eta \rangle_A~_A\langle \eta| \nu_{\lambda}
(\,d \eta)$ is constructed by this normal extension.\par
Moreover a normal extension of the $\su(1,1)$ annihilation operator is
given by the triplet 
$\left({\cal H}_{\lambda-1},|0 \rangle_{N'},
\tilde{a} \right)$,
where $\tilde{a}$ is defined as
\begin{align*}
\tilde{a} 
:= &
\left(
(A - i )\otimes \Id'+ E_+^{-1} \otimes \left( 
-\frac{1}{2} \left(L'_0 - (\lambda-1)\right) + L'_+\right) \right) \\
& \cdot
\left((A + i ) \otimes \Id'+ E_+^{-1} \otimes \left( 
-\frac{1}{2} \left(L'_0 - (\lambda-1)\right) + L'_+\right) \right)^{-1}.
\end{align*}
Next, we consider the case of $0 \,< \lambda \,< 1$.
In this case, we can similarly prove that
the triplet $\left({\cal H}_{1-\lambda},|0 \rangle_{N'},
A^* \otimes \Id'+ E_+^{-1} \otimes \left( 
-\frac{1}{2} \left(L'_0 - (1-\lambda)\right) + L'_+
\right)\right)$
is a normal extension of $A^*$.\par
Now we apply these normal extensions to a simultaneous measurement of 
$(Q^{-1}P + P Q^{-1})/2$ and $Q^{-2}$ in the boson fock space 
$L^2(\mathbb{R})$.
Since $PQ^{-1}= (Q^{-1}P + P Q^{-1})/2 + i Q^{-2}$,
we construct normal extension of $PQ^{-1}$.
In the even functions space $L^2_{{\rm even}}$
we have $A^*= PQ^{-1}, \quad \lambda = 1/2$, 
and in the odd functions space $L^2_{{\rm odd}}$ 
we have $A= PQ^{-1}, \quad \lambda = 3/2$.
Therefore, we can show that the triplet 
$\left({\cal H}_{1/2},|0 \rangle_{N'},
PQ^{-1} \otimes \Id'+ E_+^{-1} \otimes \left( 
-\frac{1}{2} \left(L'_0 - 1/2\right) + L'_+
\right)\right)$ is a normal extension of $PQ^{-1}$ in $L^2(\mathbb{R})$.
Let ${\cal H}_{1/2}$ be the even functions space in $L^2(\mathbb{R})$, 
then the normal extension of $PQ^{-1}$ is
rewritten by
$\left(L^2, | 0 \rangle_{n_b}, PQ^{-1} \otimes \Id' 
-\frac{i}{4}Q^{-2} \otimes\left( n_b' + ({a_b'}^*)^2 \right) \right)$.
\SE{Conclusions} 
We have investigated the squeezed states with various squeezing parameters 
from the viewpoint of the eigenvector system of a kind of annihilation 
operator related to the algebra $\su(1,1)$. It turned out that the 
annihilation and creation relations for this have non-linear properties, 
which is different from those of the boson case. By utilizing these 
algebraic structures, we derive an optimal POVM for the simultaneous 
measurement for $Q^{-1}P + P Q^{-1}$ and $Q^{-2}$.
We will continue more investigations of wider class of similar 
eigenvector systems based on this type of algebra.\par
\vspace{3ex}
\noindent{\large\bf References}\par
[1] H.P.Yuen, Phys. Rev. {\bf A13}, 2226 (1976) \par
[2] A.Perelomov, 'Generalized Coherent States and Their Applications',
 Springer (1986). \par
[3] I.Daubechies, 'Ten Lectures on wavelets', SIAM (1992). \par
[4] I.Daubechies \& T.Paul, Proc. 8th. Int. Cogr. Math. Phys., 
Marseilles, 675 (1986). \par
[5] F.Sakaguchi, Proc. IEEE-SP Int. Symp. Time-Frequency \& 
Time-Scale Anal., Pittsburgh, 133 (1998). \par
[6] R.Howe \& E.C.Tan, 'Non-Abelian Harmonic Analysis', Springer. \par
[7] A.S.Holevo, 'Prtobabilisitic and Statistical aspects of Quantum Theory', 
North-Holland (1982).\par
[8] M.Hayashi, 'On simultaneous measurement for non-commutative observables'
to appear in RIMS Koukyuroku, Kyoto University 
(In Japanese).\par
\end{document}